# Optimized Solar Photovoltaic Generation in a Real Local Distribution Network

Hamidreza Sadeghian, Mir Hadi Athari, Zhifang Wang
Department of Electrical and Computer Engineering
Virginia Commonwealth University
Richmond, VA, USA
Email: {sadeghianh}, {atharih}, {zfwang}@vcu.edu

*Abstract*— **Remarkable penetration of renewable energy in electric networks, despite its valuable opportunities, such as power loss reduction and loadability improvements, has raised concerns for system operators. Such huge penetration can lead to a violation of the grid requirements, such as voltage and current limits and reverse power flow. Optimal placement and sizing of Distributed Generation (DG) are one of the best ways to strengthen the efficiency of the power systems. This paper builds a simulation model for the local distribution network based on obtained load profiles, GIS information, solar insolation, feeder and voltage settings, and define the optimization problem of solar PVDG installation to determine the optimal siting and sizing for different penetration levels with different objective functions. The objective functions include voltage profile improvement and energy loss minimization and the considered constraints include the physical distribution network constraints (AC power flow), the PV capacity constraint, and the voltage and reverse power flow constraints.**

## I. Introduction

Over the past decade, employment of solar power for electricity generation has grown dramatically due to its economic benefits. However, due to variable nature of PV generation, the integration of a large amount of PV in a close geographic region will have various negative effects on the operation of distribution feeders. The most common potential concerns caused by solar power are steady-state overvoltage, impacts on system losses, and issues with voltage regulating devices, protection, and voltage fluctuation. Generally, power loss minimization and voltage stability improvement are important areas of power systems due to existing transmission financial loss of utility, network reliability, and power system blackouts. Therefore, optimal allocation of PV generation is necessary to support grid voltage regulation and improve the performance of distribution networks [1]–[4].

For the PV distributed generation (PVDG), there have been numerous studies to achieve the optimum allocation of the system. As it mentioned, the optimal site and size of DG reflects the maximum loss reduction and improvement in voltage profile of distribution system. Different methodologies have been developed to determine the optimum location and optimum size of the DG. These methodologies are either based on analytical tools or on optimization programming methods. In [5], the authors presented an analytical approach to determine the optimal allocation for the DG with an objective of loss minimization for distribution and transmission networks. In [6], a new meta-heuristic Fireworks Algorithm (FWA) is implemented on network reconfiguration problem to minimize the power loss and enhance the voltage profile of the system. A simple search algorithm is proposed in [7] for optimal sizing and placement of DG for a network system, based on losses and cost function as an objective function. The method is simple but time-consuming for searching both the best location and optimum size. In [8], the author considered the minimum loss and generation cost as a parameter in addition to DG power limits to determine the optimal size and location of the DG. The method is accurate but very tedious and mathematical computation needs much computation time. Authors in [9], presented an effective technique for optimal placement of Photovoltaic (PV) array and network reconfiguration in radial distribution network simultaneously to diminish the total real power loss and enhance the voltage level of the network. The population-based Differential Evaluation (DE) Algorithm has been implemented to identify an optimal switching combination, optimal location, size and a number of PV module in the distribution system. In [10], a stochastic approach based on kernel density estimation is proposed to identify the optimal location for the PV plant installation in distribution systems so that the voltage deviation and network losses are minimized. In order to demonstrate the effectiveness of the proposed method, the model of a real distribution feeder has been used. The test case system is located in Walterboro, SC, USA, which is composed of 38-bus and includes a photovoltaic plant. Authors in [11], have proposed a new approach to studying the impact of high PV penetration on a distribution network and its hosting capacity. The proposed method combines high-resolution resource assessment using sky imagery with power system simulation on real distribution models to study the impacts of up to 200% PV penetration level on voltage excursions, line losses, and tap changing operations. A new dual-index-based analytical approach to determine the optimal location, size and power factor of DG unit for reducing power losses and enhancing loadability is presented in [12]. This index is defined as a combination of active and reactive power loss indices by optimally assigning a weight to each index. DG placement with optimal power factor and appropriate weights of active and reactive power losses can significantly reduce losses and better enhance loadability and voltage profiles.

To conclude, there has been a lack of research on PVDG allocation based on large-scale real-world feeders that incorporates real-time solar insolation data along with time-series analysis in the U.S. and this paper aims to address this gap. The main contributions of this study are a) to build a simulation model for the local distribution network based on obtained load profiles, GIS information, solar insolation, feeder and voltage settings, b) to define the optimization problem of solar PVDG installation to determine the optimal siting and sizing and c) to analyze the optimal siting/sizing results obtained from optimization procedure. Note that the objective functions include voltage profile improvement and energy loss minimization and the considered constraints include the physical distribution network constraints (AC power flow), the PV capacity constraint, and the voltage and reverse power flow constraints.

The rest of the paper is organized as follows. In section II the description of data analysis for electrical network modeling and simulation is presented. Section III describes the problem formulation. Simulation scenarios and results are provided in section IV and section V discusses the results and concludes the paper.

## II. DATA ANALYSIS AND ELECTRICAL NETWORK MODELING

The hypothetical electrical distribution model is based on a local utility distribution network in Virginia with a summer peak of 23,260 kW feeding 1902 customers by classes of 1429 residential, 397 small commercial, and 76 large commercial/industrial. The network voltage level is 12.5 kV. For security reasons, the local utility could not provide specific information on electrical network structure and design. Therefore, we tried to model the system based on the logical alignment of the electrical system and statistical analysis of available data from the local utility and Open Energy Information (OpenEI).

By obtaining GIS layers that identified structures and zoning information for all properties in the area surrounding the substation and an iterative re-drawing process, a collection of buildings around the substation that perfectly matched the local utility customer data for the substation captured. For electrical network modeling, first, study area divided to six sub-regions based on study area map and electrical lines and then main distribution feeders and branch lines which recognized from Google map, logically assumed to feed these areas. An exact number of building types in each area derived from GIS layers. (Fig. 1).

### A. Load buses and load profiles

In order to consider load buses, buildings in the study area are grouped based on geographical locations. Next, load profile of each bus is calculated by aggregating of hourly load profiles of all buildings connected to that bus [13]. Note that, different bus types are determined based on connected building types.

OpenEI dataset includes hourly load profiles for all types of buildings (e.g., residential base load, residential low load, large office, small office, quick service restaurant, small hotel, etc.) during a year. The peak load for each load type is determined based on statistical analysis of available realistic data. Fig. 2 shows the empirical probability density function (PDF) of peak load for base residential load and the approximated fit distribution. The empirical PDF is approximated with Log-Logistic distribution with parameters shown in Fig. 2.

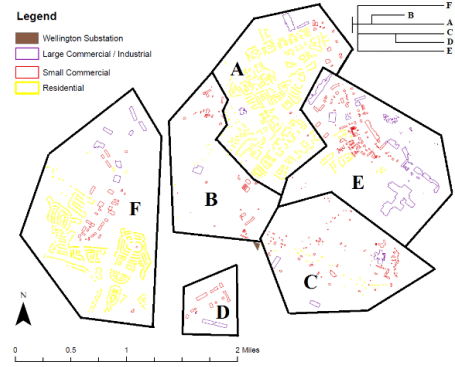

Figure 1. Base map of study area and sub-regions

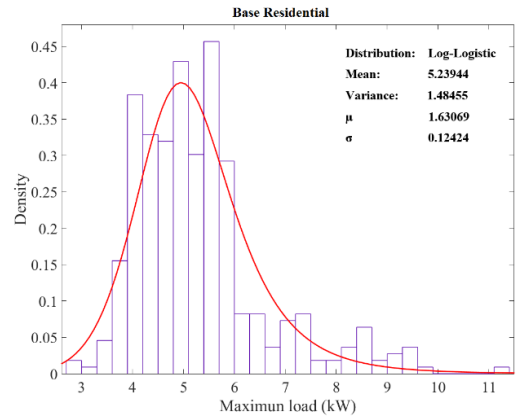

Figure 2. Empirical PDF of base residential load and approximated fit distribution

By considering provided actual peak load by the local utility and scaling bus load profiles, aggregated load profiles for all categories are shown in Fig. 3.

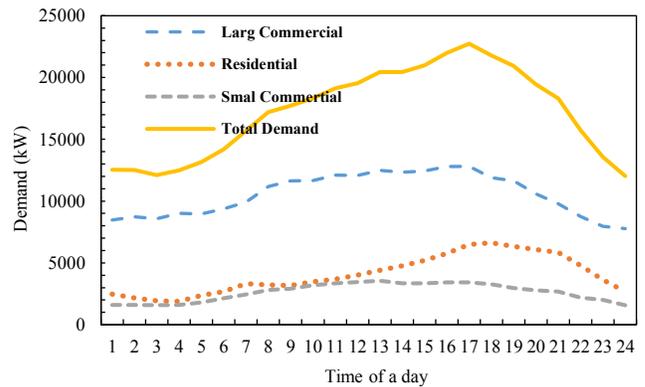

Figure 3. Aggregated load profiles for all categories

## B. Solar radiation potential

Solar insolation was determined using LiDAR elevation source data from the US Geological Survey (USGS) "The National Map" (TNM) Download Manager service (U.S. Geological Survey, 2016). This data comes in the form of CVS point files, which can then be converted into a raster file in ArcGIS. Once the LiDAR data is in raster form, the "Area Solar Radiation" tool in ArcGIS can convert that data into a solar insolation raster, as shown in below in Fig. 4. Given that these raster values stem from the LiDAR data, they take into account shading from trees and other obstructions. For buildings with pitched roofs, the process also reflects that south-facing surfaces receive higher average annual insolation. The solar insolation raster data can then be converted back into point data, showing the average annual solar insolation for a given area, in terms of watt-hours of solar radiation per year.

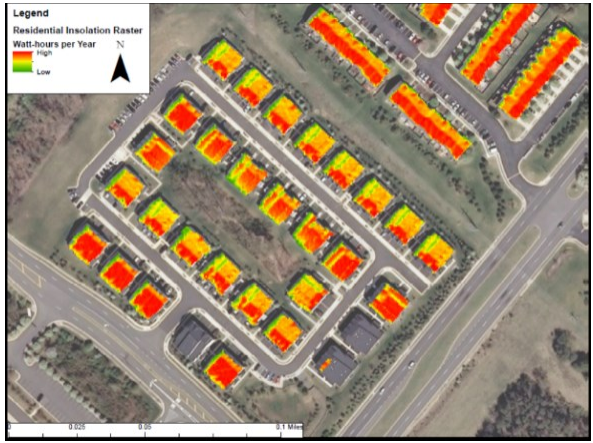

Figure 4. Solar insolation raster values in study Area

At this point, the average insolation could be calculated for each building. However, simply calculating the average insolation for an entire building surface would under-estimate the potential of a PV array placed on that building, as in the real world PV arrays are only placed in optimal locations that will receive adequate insulation to maximize their cost-effectiveness. Therefore, a query applied so that the insolation points layer would only display points with an average solar insolation of greater than 4.6 kWh/m$^2$/day. This threshold value produced clearly discernable patterns of high-insolation points. Next, the percentage of each building that is covered with those high-insolation points is calculated, and divided the buildings into three categories based on their density of high-insolation coverage, as shown in Table 1. Figure 5 shows the distribution of low, medium, and high-insolation buildings (defined by the density of high-insolation points on each roof) within a selected neighborhood of single-family attached housing. The map demonstrates that the buildings with minimal shading and east-west orientation (i.e., those with south-facing rooftops) are more likely to be designated as high-insolation in this methodology.

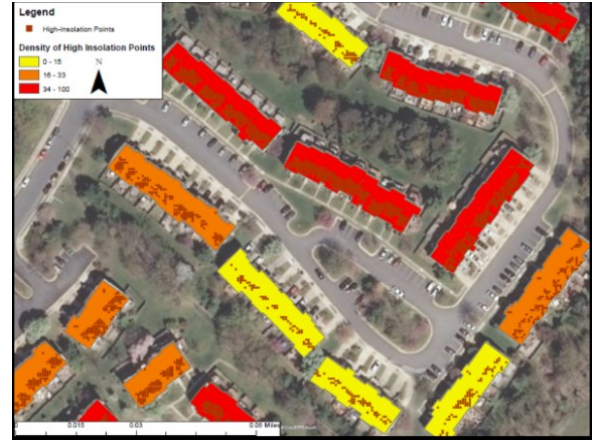

Figure 5. Distribution of low, medium, and high Insolation on buildings

The final steps of this part were to identify the potential amount of distributed PV that could be installed in the study area and the ensuing amount of potential electricity production. We first assumed that PV would only be installed buildings that fall into the "high-insolation" category. This meant that 510 of the residential buildings (36%), 119 of the small commercial buildings (30%), and 26 of the large commercial and industrial buildings (34%) would be eligible for PV installations. Aggregating the values per building to determine the total potential installed solar power capacity and electricity generation in the study area, by building type category shows that that the total potential PV coverage of 24,830 kW is equal to 107% of the total sub-station area peak load. The potential annual energy production of 32,310 MWh is equal to 28% of the estimated annual energy demand of 114,758 MWh for the entire substation service area. These figures indicate that a substantial portion of the study area's electricity demands could be met through local distributed solar power, assuming that distribution grid impacts could be mitigated. By considering high-insolation buildings we can assume buses with the potential to install PV systems.

TABLE I. PERCENT OF BUILDING COVERAGE FOR ALL CATEGORIES

| Building Type | Percent of Building Coverage with Insolation above 4.6 kWh/m$^2$/day | | |
|---|---|---|---|
| | Low | Medium | High |
| Residential | 0 – 15% | 16 – 33% | 34% or above |
| Small Commercial | 0 – 15% | 16 – 33% | 34% or above |
| Large Commercial / Industrial | 0 – 25% | 26 – 40% | 41% or above |

## C. Electrical parameters

For calculating reactive power profile for each bus, statistical evaluation of power factor for loads from given sub-station hourly data is performed. It is found that the power factor for aggregated loads follows a normal distribution with μ=0.8233 and σ= 0.0193. Next, given the active demanded power for each load, the reactive power profile is calculated according to random power factors generated from a normal distribution. For the electrical network, line parameters are calculated based on the estimated distance between buses, and conductor electrical characteristics. The schematic diagram of the electric networks is shown in Fig.6.

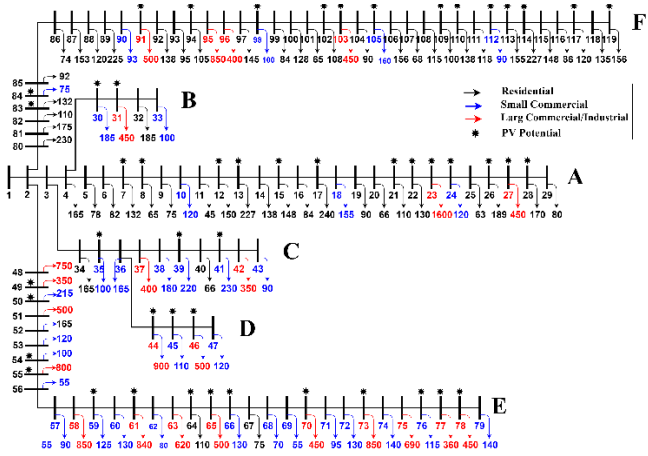

Figure 6. Schematic diagram of distribution network

## III. PROBLEM FORMULATION

The main objective of the proposed method is to determine the optimal placement and sizing of PV systems that minimizes the multi-objective function of power loss and voltage deviation subject to distributed PV constraints and operational constraints of a distribution network such as avoiding reverse power flows in the network. PV penetration ratio is defined based on substation peak load and is as follows

$$\alpha(\%) = \frac{\sum_{i=1}^{N} P_{PV_i}^{max}}{\max\left(\sum_{j=1}^{M} P_{Load_j}\right)} * 100 \quad (1)$$

where $P_{PV}$ and $P_{Load}$ are PV panel output power (kW) and electrical load demand (kW), respectively. In this paper penetration ratios are considered as 5, 10, 20, and 50%.

The total real energy loss of radial distribution system can be calculated as [8]

$$E_{loss} = \sum_{t=1}^{24} \sum_{l=1}^{L} |i_L^t|^2 R_L \quad (2)$$

where $i_L^t$ is current flowing through line $L$ at time $t$ and $R_L$ is resistance of line $L$. The formulation for voltage profile improvement ($v_D$) with $v_n^t$ as the voltage of bus $n$ at time $t$ is as follow [14]:

$$v_D = \sum_{t=1}^{24} \sum_{l=1}^{n} (v_i^t - 1)^2 \quad (3)$$

The multi-objective function can be formulated as follows, the minimum of the objective function implies the best PV-DG allocation for minimizing power losses and enhancing loadability and voltage profiles.

$$\min_{\mathcal{L}_{PV}, P_{PV}^{max}} f = E_{loss} + \omega * v_D \quad (4)$$

Subject to:
$$f(P_L, P_{PV}, v|Y_{bus}) = 0 \quad (5)$$
$$i = h(v|Y_{bus}) \quad (6)$$
$$P_{PV}(t) = g(P_{PV}^{max}, I(t)) \quad (7)$$
$$0 \leq P_i^{max} \leq \widehat{P_{PV,i}^{max}} \quad (8)$$
$$1^T P_{PV}^{max} \leq \alpha. P_{sub}^{max} \quad (9)$$
$$i_{ij}^t \geq 0 \quad \forall \, i < j \quad (10)$$

$$0.95 \leq |v_i| \leq 1.05 \quad (11)$$

where $\mathcal{L}_{PV} = [\ell_1, \ell_2, ..., \ell_n]^T$ $\ell_i \in \{0,1\}$ is PVDG location vector, ω is weighting factor, $P_{PV}^{max} = [P_1^{PV}, P_2^{PV}, ..., P_n^{PV}]$ is PVDG maximum capacity vector, $Y_{bus}$ is network admittance matrix, $i$ is vector of bus injected current, $v$ is bus voltage vector, $I(t)$ is solar insolation at time $t$, $\widehat{P_{PV,l}^{max}}$ is PV installation limit for bus $i$ derived from solar data analysis, and $i_{ij}^t$ is current flowing from bus $i$ to $j$ at time $t$, $L$ is the number of lines, $n$ is the total number of buses, and $P_{sub}^{max}$ is defined as substation peak load which is 23,260 kW in the experimented case. Note that Eqs. (5) and (6) are network constraints enforced by AC power flow and network operation constraints, respectively. Eqs. (7-9) are PVDG installation constraints.

With consideration for the effect of temperature, the output generated power (kW) by PV arrays is determined using the model from [15]. Note that in this study, both direct and diffuse beam irradiance data are considered in the calculation of PV array output.

## IV. SIMULATION SCENARIOS AND RESULTS

Four scenarios based on different penetration ratios of PV are considered to study the impacts of PV on voltage and current profiles, network power loss, and reverse power flow. The Improved Particle Swarm Optimization (IPSO) is used to solve the objective functions of optimum PVDG placement and sizing [15].

All the simulations and load profiles carried out for June 24 base on the actual occurrence time of maximum substation load. It should be mentioned that for this day, total system energy consumption is 421.22 (MWh) and total energy loss is calculated 26.99 (MWh) which is 6.40% of total system energy consumption. Desired system operation voltage is set to 1 p.u., so for modeled system, the voltage deviation (VD) is calculated as 1.8315. All solar potential buses are a candidate to have PV installation and the installation capacity limit is proportional to calculated solar power potential in data analysis section. Also, in this way the physical limitation on rooftop area for buildings is taken into account. In order to apply different objectives of PVDG systems implementation, all possible objective functions are investigated in the simulation section. Therefore, in the comparative performance study, four individual objectives are considered: I) Voltage improvement and energy loss reduction with reverse power flow constraint. II) Voltage improvement and energy loss reduction. III) Energy loss reduction VI) Voltage improvement. The total energy loss and voltage improvements for different penetration levels for all objective functions are presented in Table II. As shown in Table II, total energy loss is decreased by the deployment of PVDG systems. However, the analysis shows that voltage deviation for the network is not a strictly decreasing function and its minimum occurs at 20% penetration ratio. This means, increasing penetration ratio of PVDGs always would not necessarily result in improved voltage profiles for distribution network and for penetration ratios above a certain threshold, there would be voltage quality and reverse power flow issues that need to be addressed. It is worthwhile to note that all the buses for PV installation are selected from PV potential buses derived in data analysis section.

There are all types of residential, small commercial and large commercial buses in the optimum solution. However, large commercial buses because of large PV installation capacity and alignment of their profile with solar irradiance profile, play an effective role in the optimum solutions. Voltage violation is considered as a constraint for all PV installation scenarios, so there is not any violation in voltage. Fig. 7 shows voltage profiles for all PV penetration ratio.

TABLE II.  TOTAL ENERGY LOSS AND VOLTAGE DEVIATION FOR ALL SCENARIOS

| α | Objective function # | Energy loss | | Voltage deviation | | Random installation |
|---|---|---|---|---|---|---|
| | | kWh | Imp. (%) | $v_D$ | Imp. (%) | |
| 5% | I | 26,081 | 3.39 | 1.7372 | 5.15 | Energy loss 26,358 |
| | II | 26,067 | 3.44 | 1.7375 | 5.13 | |
| | III | 25,996 | 3.70 | 1.7325 | 5.41 | $v_D$ 1.7688 |
| | VI | 26,083 | 3.38 | 1.7250 | 5.81 | |
| 10% | I | 25,837 | 4.29 | 1.6759 | 8.50 | Energy loss 25,914 |
| | II | 25,216 | 6.59 | 1.6701 | 8.81 | |
| | III | 25,233 | 6.53 | 1.6711 | 8.76 | $v_D$ 1.6927 |
| | VI | 25,289 | 6.32 | 1.6628 | 9.21 | |
| 20% | I | 24,092 | 10.76 | 1.6102 | 12.08 | Energy loss 25,187 |
| | II | 24,167 | 10.48 | 1.6192 | 11.59 | |
| | III | 23,798 | 11.85 | 1.5858 | 13.42 | $v_D$ 1.6494 |
| | VI | 24,001 | 11.09 | 1.5848 | 13.47 | |
| 50% | I | 22,998 | 14.81 | 1.6295 | 11.03 | Energy loss 23,338 |
| | II | 22.268 | 17.51 | 1.6218 | 11.45 | |
| | III | 22,131 | 18.02 | 1.6228 | 11.40 | $v_D$ 1.6306 |
| | VI | 22,428 | 16.92 | 1.6206 | 11.52 | |

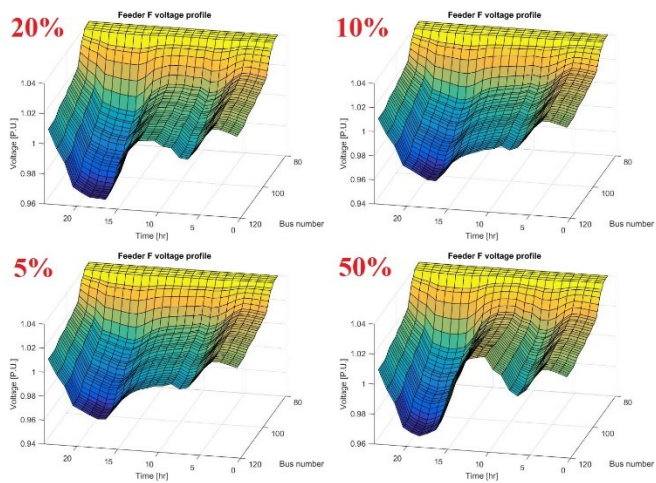

Figure 7.  Voltage profile of feeder F for all scenario

## V. CONCLUSION AND RECOMMENDATIONS

Optimal allocation of PVDG with different objective functions including voltage profile improvement and energy loss reduction is presented in this paper. The optimization problem is solved considering both individual objective function and multi-objective functions. One of the greatest challenges to the insertion of distributed generation, especially to the use of photovoltaic technology, is the utilization of its benefits without losses in reliability and with satisfactory operation of electrical power systems. In this context, voltage profile along the feeders and magnitude and direction of current through lines are of great importance. Therefore, The main contributions of this study is building a simulation model for the local distribution network based on obtained load profiles, GIS information, solar insolation, feeder and voltage settings, and define the optimization problem of solar PVDG installation to determine the optimal siting and sizing . Optimum solutions is found for different penetration levels with different objective functions.  As the future extension of this study we can consider annual analysis along with economic analysis to suggest best strategy for PVDG allocation.

REFERENCES

[1] M. M. Haque and P. Wolfs, "A review of high PV penetrations in LV distribution networks: Present status, impacts and mitigation measures," *Renew. Sustain. Energy Rev.*, vol. 62, pp. 1195–1208, 2016.
[2] M. J. E. Alam, K. M. Muttaqi, and D. Sutanto, "A comprehensive assessment tool for solar PV impacts on low voltage three phase distribution networks," *2012 2nd Int. Conf. Dev. Renew. Energy Technol.*, pp. 1–5, 2012.
[3] G. Guerra, J. a Martinez, and S. Member, "A Monte Carlo Method for Optimum Placement of Photovoltaic Generation Using a Multicore Computing Environment," *PES Gen. Meet. Conf. Expo. 2014 IEEE. IEEE,* pp. 1–5, 2014.
[4] T. Tayjasanant and V. Hengsritawat, "Comparative evaluation of DG and PV-DG capacity allocation in a distribution system," *Proc. Int. Conf. Harmon. Qual. Power, ICHQP*, pp. 293–298, 2012.
[5] C. Wang and M. H. Nehrir, "Analytical Approaches for Optimal Placement of Distributed Generation Sources in Power Systems," *IEEE Trans. Power Syst.*, vol. 19, no. 4, pp. 2068–2076, Nov.2004.
[6] A. Mohamed Imran and M. Kowsalya, "A new power system reconfiguration scheme for power loss minimization and voltage profile enhancement using Fireworks Algorithm," *Int. J. Electr. Power Energy Syst.*, vol. 62, pp. 312–322, 2014.
[7] S. Ghosh, S. P. Ghoshal, and S. Ghosh, "Optimal sizing and placement of distributed generation in a network system," *Int. J. Electr. Power Energy Syst.*, vol. 32, no. 8, pp. 849–856, 2010.
[8] A. Medina, J. C. Hernández, and F. Jurado, "Optimal placement and sizing procedure for PV systems on radial distribution systems," *2006 Int. Conf. Power Syst. Technol. POWERCON2006*, 2007.
[9] M. Dixit, P. Kundu, and H. R. Jariwala, "Optimal Placement of PV Array in Distribution System for Power Loss Minimization Considering Feeder Reconfiguration," 2016.
[10] Y. Chen, M. Strothers, and A. Benigni, "A Stochastic Approach to Optimum Placement of Photovoltaic Generation in Distribution Feeder," 2016.
[11] A. Nguyen *et al.*, "High PV penetration impacts on five local distribution networks using high resolution solar resource assessment with sky imager and quasi-steady state distribution system simulations," *Sol. Energy*, vol. 132, pp. 221–235, 2016.
[12] D. Q. Hung and N. Mithulananthan, "Loss reduction and loadability enhancement with DG: A dual-index analytical approach," *Appl. Energy*, vol. 115, pp. 233–241, 2014.
[13] "Index of /datasets/files/961/pub." [Online]. Available: http://en.openei.org/datasets/files/961/pub/.
[14] S. Kucuksari *et al.*, "An Integrated GIS, optimization and simulation framework for optimal PV size and location in campus area environments," *Appl. Energy*, vol. 113, pp. 1601–1613, 2014.
[15] M. H. Athari and M. M. Ardehali, "Operational performance of energy storage as function of electricity prices for on-grid hybrid renewable energy system by optimized fuzzy logic controller," *Renew. Energy*, vol. 85, pp. 890–902, 2016.